# Metastable state involved resonant tunneling through single InAs/GaAs quantum dot

*TianXin Li, PingPing Chen, DaYuan Xiong, ZhaoLin Liu, XuMing Zhang, ChangSheng Xia, ZhiFeng Li, XiaoShuang Chen and Wei Lu[*]*

National Laboratory for Infrared Physics, Shanghai Institute of Technical Physics, Chinese Academy of Sciences, 500 YuTian Road, Shanghai 200083, People's Republic of China

Corresponding author: luwei@mail.sitp.ac.cn

Abstract

A scheme of resonant tunneling through the metastable state of semiconductor quantum dot is presented and implemented in the transport study of freestanding InAs quantum dots grown on GaAs(001) under illumination using conductive atomic force microscopy. The metastable state is achieved by capturing one photoexcited Fermi hole in the valence energy level of InAs quantum dot. Resonant tunneling through single quantum dot can be observed at room temperature due to the existence of metastable state. The amplitude of tunneling current depends on the barrier arrangement and the concentration of photoexcited holes around the quantum dot, but is found steady when the height of dot varies from 1.8 to 9.9 nm, which are in good agreement with the proposed model. The experiment demonstrates a solution of room temperature operated single electron device to amplify the photocurrent by the singularity of resonant tunneling in epitaxial quantum dot.

Confined system with discrete energy spectrum is a promising site to monitor and control the storage,[1,2] motion[3,4] and transition[5,6] of single electron, which might provide great potential applications for future quantum information technology. Among those practical candidates of atomlike electronic structure, epitaxially self-assembled semiconductors in nanometer-size, commonly called quantum dots (QD),[7] have the advantage to realize the opto-electronic utility based on single electron process. According to Zrenner *et al.*'s experiment,[8] the QDs can work as 'turnstile' devices for sequential electrons excited by pulsed laser. And the potential change induced by single charge occupancy in a QD has been used for changing the transport status of a resonant tunneling diode (RTD) which leads to efficient single photon detection.[9] So far the single electron characters in individual self-assembled QD were mostly demonstrated at ultra low temperatures in order to resolve the small energy scales.[10,11,12,13] It remains a big challenge to keep the singularity of resonant tunneling in single QD at room temperature which is critical for it's application in the electrical or opto-electronic amplifying process. On the other hand, single electron tunneling was mainly accomplished in static bound states system up to now; few efforts have been focused on resonant tunneling through metastable state of self-assembled QD, though similar mechanism has been discovered in lithography-defined QD and opens new access of manipulating the quantum transport process in confined systems.[14]

Normally, for a double-barrier isolated QD with mean energy level spacing $\Delta$, the thermal smearing effect will prevent us from distinguishing the sublevels of QD during tunneling measurement when the thermal energy $kT$ is close to $\Delta$ (Fig 1a).[15] The scheme involved in this study is related to resonant tunnel of electrons through the metastable state (an individual valence energy level) of InAs QD under particular band setup, namely this "valence band (VB)" level is aligned with the conduction-band edge of GaAs collector (Fig 1b). The metastable state is achieved by filling the QD with one photoexcited hole. At the same time, the neighboring "VB" levels, which are filled with electrons, will not participate in the tunneling process due to the Pauli principle for Fermi carriers. Consequently, taking advantage of the metastable hole state in QD, the single level resonant conductance in a double-barrier quantum dot system can be maintained even at room temperature.

The InAs/GaAs samples are grown on GaAs(001) substrates by molecular beam epitaxy (MBE). A 1200-nm-thick Si-doped ($1\times10^{18}$ cm$^{-3}$) GaAs is first deposited as the bottom electrode layer followed by 4 monolayers (MLs) of GaAs spacer, the QDs is then self-assembled by depositing near 2.4 MLs of InAs at relevant temperatures (500 ~ 520 °C) for dot-size variation (sample *a*, Fig 1c). The energy band bending near the surface of *n*-GaAs constitutes a barrier potential separating the dots with the collector. Samples with 5-nm-thick Al$_{0.3}$Ga$_{0.7}$As barrier layer between the QDs and the *n*-GaAs (sample *b*), as well as the dots on *p*-type doped GaAs (sample *c*), are also prepared for comparison.

The transport properties of InAs QDs are studied at room temperature by a MultiMode atomic force microscope (MMAFM, Veeco instruments). The measurements are done in dry air (RH < 20%) or nitrogen ambient in order to minimize the influence of the moisture on sample surface. A tunneling amplifier with sensitivity of 10 pA/V is used to detect the tip to sample current in contact mode. Silicon tips coated with PtIr alloy film are used for both topography and local current measurements. The native oxide of air-exposed InAs surface makes of a narrow interfacial barrier between the metal tip and the semiconductor. Then, as shown in Fig 1c, the tip-dot-substrate setup constitutes a double-barrier tunnel junction; The Fermi level of GaAs substrate is deeply pinned in the bandgap due to the large surface state density of InAs polar facet.[16,17]

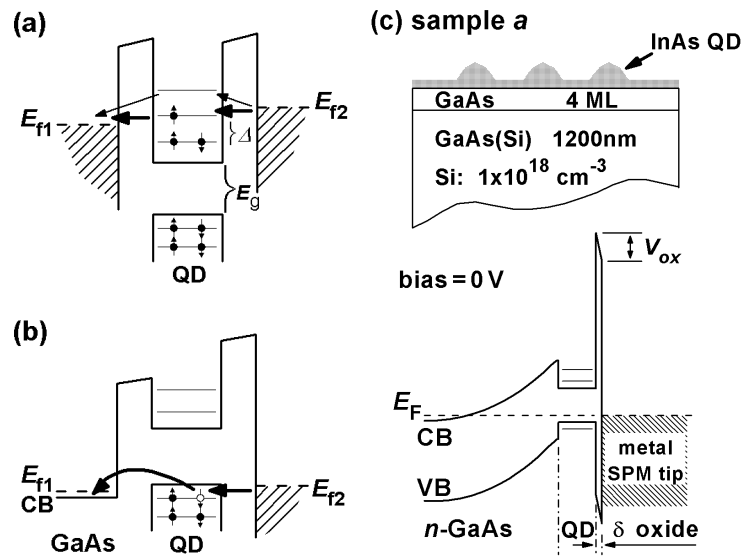

**Figure 1**. Schemes representing the principles of tunneling of electrons through (a) the conduction energy levels of QD when $kT \sim \Delta$, (b) the metastable hole state (presented as white circle) in "valence

band" of QD; (c) Structure of sample *a* (up) and the band profile of Conductive-AFM measurement on freestanding InAs QD under zero bias voltage (bottom). $\Delta$: mean sublevel spacing in QD, $V_{ox}$: discontinuity of vacuum level induced by the surface states of InAs/GaAs.

Repeatable electrical measurements on individual QD in air usually need careful selection of the parameters. First, the lateral drift of SPM tip shall be essentially suppressed (less than 5 nm per minute), since the detected conductivity of a QD is likely to be inconsistent when the tip is positioned at different points of the dot.[18,19] Second, the repulsive force between the SPM tip and the QD should be delicately adjusted to eliminate the instability of contact resistance. In our case, this can be achieved by gradually increasing the repulsive force to a few tens of nano-newton.

The inset of figure 2a presents a height and current mapping simultaneously obtained on sample *a* under a sample to tip bias of 230 mV. Current of a few picoamperes through the InAs QDs can be measured while the wetting layer remains resistive. This is consistent with previous reports about the less conductivity of wetting layer compared with freestanding QDs.[19,20]

Figure 2 also shows the *I-V* curves of individual QDs on variant substrates. The plots in figure 2b represent electrical transport behavior of the dots on *p*-type GaAs substrate with different sizes (sample *c*), which can be related to the commonly known *I-V* property of a Schottky contact. However, for quantum dots on *n*-type doped substrate (sample *a* and *b*), both curves in figure 2a exhibit step-like structures under positive sample to tip bias. Similar phenomenon has been also found on InGaAs/*n*-GaAs(311) QDs at relatively low temperature.[19] Here we will demonstrate that this distinct character is associated with the scheme of resonant tunneling through the hole occupied "VB" level of QD.

First, it can be disclosed that the current steps in figure 2a are originated from the photoexcited carriers. The excitation light is a fraction of the diode laser beam which is provided in the AFM head for cantilever deflection detection (690 nm, 1.0 mW in our case). The irradiation mainly occurs around the tip engaged location of sample surface when the width of cantilever is smaller than the beam diameter (We use the probes with cantilever of 20 μm in width and is triangular at the front end). By reducing the power density of incident diode laser on tip-engaged area, the step height in the *I-V* curve of QD on sample *a* is found essentially diminished (see the black plots in the inset of figure 3). An extra light from Ti:sapphire laser (905 nm) is then introduced to the detecting area. One can find in figure 3 that, within the applied illumination power, the step height of current increases steadily with the excitation intensity.[21] This dependence of current step on the light intensity confirms that the step like structure in figure 2a is induced by photoexcited carriers.

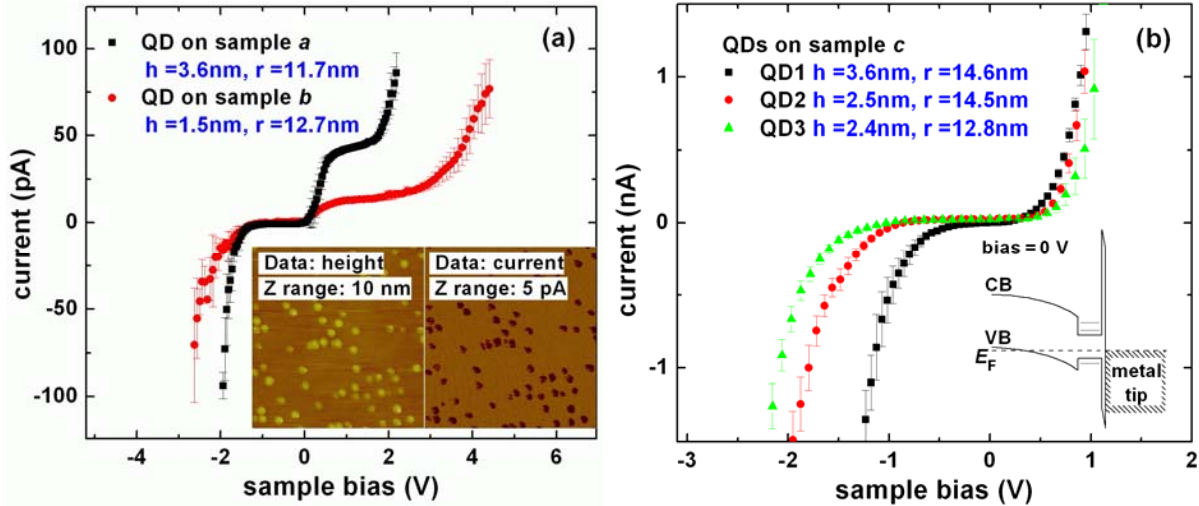

**Figure 2**. *I-V* curves measured on individual InAs QDs of sample *a*, *b* (a) and sample *c* (b) respectively with the size labeled by height (h) and base radius (r) of the dot. The insets in (a) are the morphology and current images obtained simultaneously on sample *a* at a sample bias of 230 mV with scan size 650 × 650 $nm^2$. The inset in (b) is the schematic band profile of tip-QD/*p*-GaAs junction under zero bias.

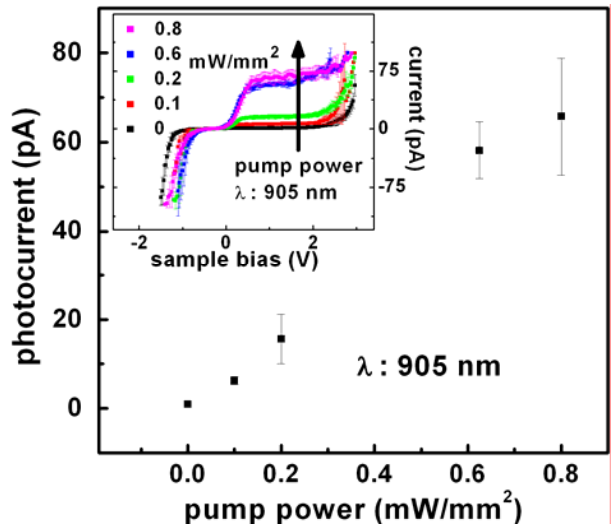

**Figure 3**. Dependence of the step height in *I-V* curves (inset) on the excitation intensity. Inset: *I-V* curves of one InAs QD on sample *a* under variant irradiation.

Secondly, the energy band alignment in the tip-sample structure is critical for the *I-V* property under illumination. As we have mentioned above, electrical property similar to that of a normal Schottky contact is observed in the QDs on *p*-type GaAs buffer within a wider extension of current. The surface potential makes the essential difference between the *n*- and *p*-doped substrates. For *n*-doped substrate, the upward bending band near surface serves as a barrier for electrons and makes the QD a trap for Fermi holes when light excitation happens. The tip-dot/*n*-GaAs setup with occupied hole state in the QD is thus a prototype of our proposed scheme of resonant tunneling shown in figure 1b. On the other hand, the Fermi level of *p*-type doped GaAs buffer layer locates close to the top of valence band (inset of figure 2), holes can tunnel through the valence levels of QD under appropriate bias.[22] So the tunneling transport of tip-dot/*p*-GaAs junction is substantively subject to the scheme illustrated in figure 1a in which thermal smearing effect can not be suppressed.

In the case of positive sample to tip bias for QDs on *n*-doped substrates, normally the tunneling current should be negligible for the existence of QD's bandgap and the strengthened surface barrier (Figure 4a). However, with increase of bias voltage, the "VB" levels in QD will be aligned from below to above the conduction-band edge of GaAs, the hole occupied state in "VB" will thus provide the channel for electrons tunneling from tip to the conduction band of GaAs substrate. Resonant tunneling through the hole occupied state of QD leads to abrupt increase of the current. This enhancement of tunneling current will be maintained, which results in a step-like structure in the *I-V* curve, before the reverse breakdown of tip-sample contact. Contrarily, a forward bias for tip-sample contact counteracts the surface barrier (Figure 4b). At certain bias, tunneling of electrons through the conduction levels of QD may happen with the assistance of thermal energy at room temperature, so that the forward breakdown is often observed in *I-V* tendency with relatively lower threshold bias than that of reverse breakdown.[20]

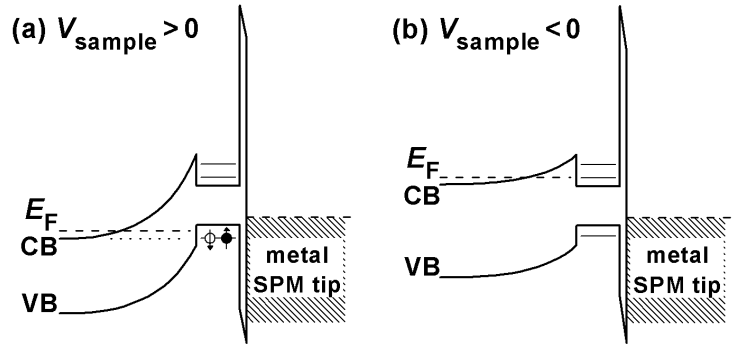

**Figure 4**. Schematic band structures of tip-QD-substrate junction (sample *a*) representing resonant tunneling of electrons through the hole state in "valence band" at positive bias (a) and the conduction levels of QD at negative bias (b).

The amplitude of resonant tunneling current, namely the step height in figure 2a, is mainly dominated by the transmission rate of double barrier assembly as well as the occupancy and recombination of holes in the QD. With similar detecting scheme based on scanning probe microscopy (SPM), resonant tunneling through individual energy level of QD usually yields a current from several to tens of picoamperes depends on the concrete spacer or barrier arrangements.[18,22,23] The maximum current of resonant tunneling we measured on the QDs of sample *a* is about 60 pA, we attribute this to the relatively high transmission rate of the tip-dot assembly as a result of the minimized width of both side's barriers of the QD. Comparatively, when the surface barrier is enhanced by $Al_{0.3}Ga_{0.7}As$ layer, the step height shown in figure 2a becomes significantly lower, ~15.4 pA, for sample *b*. Another major factor that influences the current amplitude is the duty ratio of metastable hole state in QD. This can be evidenced by the consecutively ascending of the step height in *I-V* curves with the increasing excitation intensity, since the duty ratio of hole occupancy in a dot is proportional to the hole concentration excited by illumination light.[24]

The dependence of transport behavior on the size of QD is finally investigated to reveal the quantum confinement effect in resonant tunneling process. Figure 5a are the *I-V* curves measured on QDs in different sizes of sample *a*. The background excitation is typically the same with that in figure 2a, and all the measurements are performed within one engagement with the same SPM tip for consistency. The height of the step in each of the *I-V* curves is around 44.5 pA and varies little with the size of QD. This implies that the number of quantum state involved in the resonant tunneling process is the same for QDs in different sizes, since the transmission rate of a double-barrier junction depends on the effective density of states rather than on the dimension of the confined structure.

However, different size of QD will result in different energy level alignment of occupied hole state in QD, the threshold voltage of the step in *I-V* curve is then supposed to be affected by the size of QD according to the model of Fig.1(b). Three threshold biases can be defined according to the inflexions in the *I-V* curves of each QD: the threshold bias for resonant tunneling through hole-occupied "VB" level of QD ($V_{RT}$), the bias for forward breakdown ($V_{FB}$) and reverse breakdown bias ($V_{RB}$). The *dI/dV* curves of a series of QDs reveal the trends of the threshold biases along with the dot size (Figure 5b).   1) For all the dots we've studied, the bias value needed for *reverse* conduction is lower than that for *forward* conduction, namely $|V_{RT}| > |V_{FB}|$. This provides further evidence that the quantized electronic levels of QD should be included in the tunneling process; and it indicates that the Fermi level of GaAs surface is pinned close to the valence band of the InAs quantum dot, which is consistent with the findings of photoreflectance studies on the InAs QDs covered GaAs(001) surface in reference 17.   2) Both $|V_{RT}|$ and $|V_{FB}|$ increase monotonically when the height of QD decreases from 9.9 nm to 1.8 nm (Figure 5b). The tendencies can be attributed to the broadening bandgap ($E_g$) with shrinking size of the QD. A comparison between the bias gap, $|V_{RT} - V_{FB}|$, and the estimated $E_g$ is shown in figure 6. The band gap energy is calculated in the framework of effective mass approximation to take the effects of dot shape, confining potential and the strain into account. The $|V_{RT} - V_{FB}|$ follows similar size-dependency with $E_g$ as expected. However, the extracted bias gaps are commonly larger than the bandgap of quantum dots, which indicates the strong blockage effect on the band bowing due to the limited size of tip-dot Schottky contact.[20]

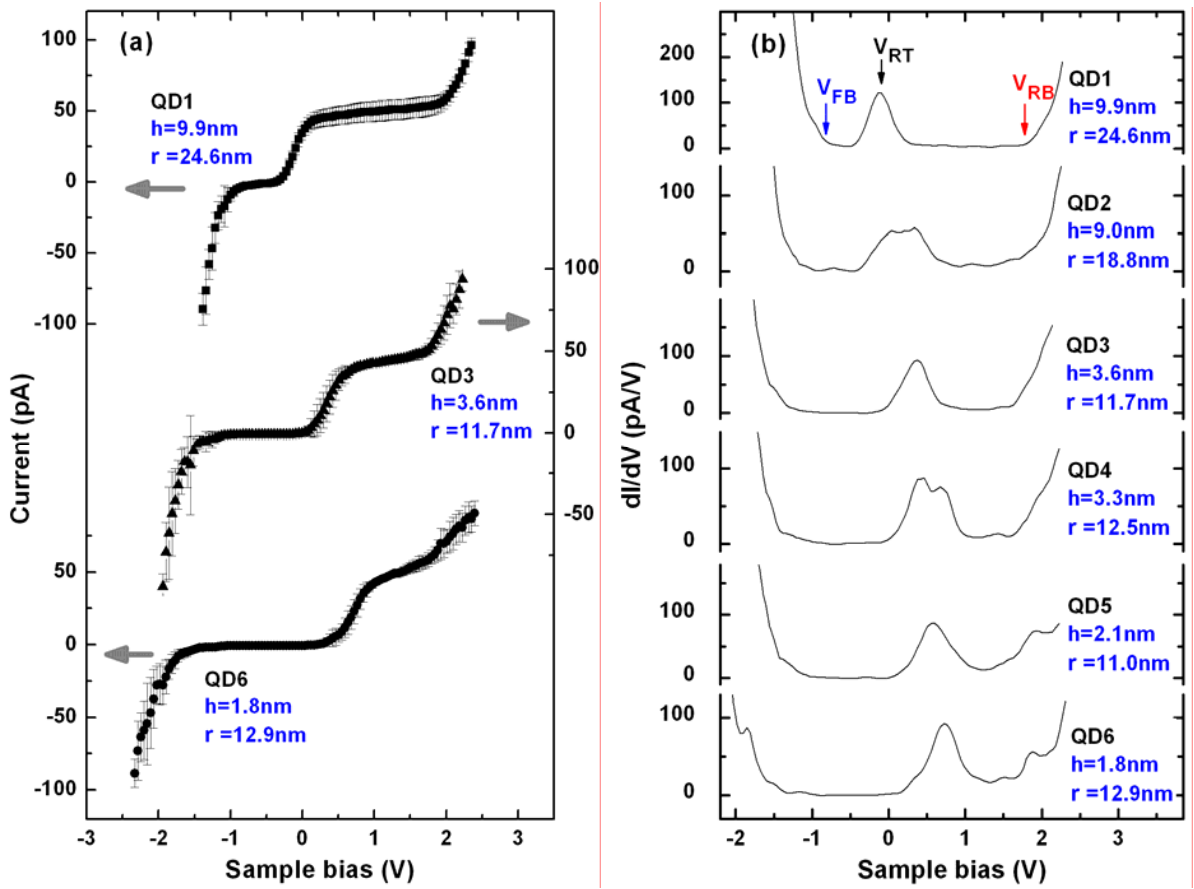

**Figure 5**. The *I-V* plots (a) and *dI/dV* curves (b) of QDs with various sizes on sample *a*. The *I-V* measurement of each dot is repeated for 5 to 10 times.

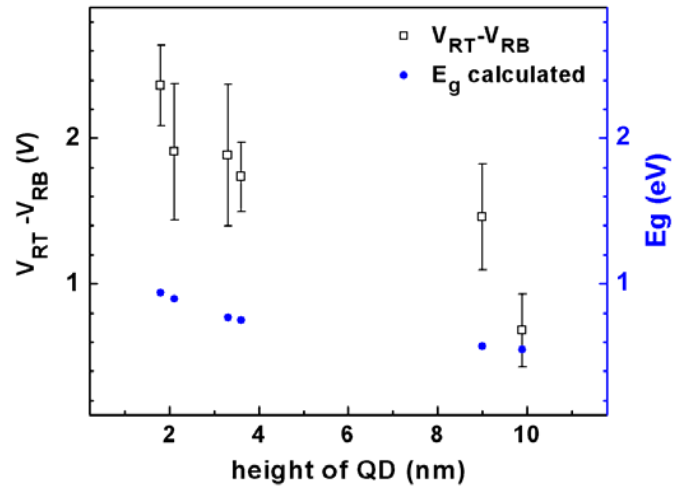

**Figure 6.** Characteristic bias gap |$V_{RT}$ - $V_{FB}$| and the estimated energy-gap ($E_g$) versus the height of QDs.

In summary, we report the observation of room temperature resonant tunneling through the hole occupied valence energy level in single InAs/*n*-GaAs quantum dot. The conductivity of tunneling junction can be changed by altering the barrier arrangement or by modulating the concentration of holes around QD, but is independent of the dot size. By using the metastable state in quantum dot, our demonstrated scheme provides a route for the novel functional opto-electronic devices based on the room temperature operating single electron process.

**Acknowledgment.** The author (T. X. Li) thanks Dr. Jun Shao for his help with laser technique. This work was financially supported by the Grand Foundation of Shanghai Science and Technology (05DJ14003), Shanghai Applied Material Foundation (0521), National Natural Science Foundation of China (10604059), State Key Basic Research Program of China (2006CB921507, 2007CB613206) and the Knowledge Innovation Program of the Chinese Academy of Sciences.

References

(1) Lundstrom, T.; Schoenfeld, W.; Lee, H.; Petroff, P. M. *Science* **1999**, *286*, 2312.

(2) Segal, D.; Shapiro, M. *Nano Lett.* **2006**, *6*, 1622.

(3) Guisinger N. P.; Greene, M. E.; Basu, R.; Baluch, A. S.; Hersam, M. C. *Nano Lett.* **2004**, *4*, 55.

(4) Bjork, M. T.; Thelander, C.; Hansen, A. E.; Jensen, L. E.; Larsson, M. W.; Wallenberg, L. R.; Samuelson, L. *Nano Lett.* **2004**, *4*, 1621.

(5) Stevenson, R. M.; Young, R. J.; Atkinson, P.; Cooper, K.; Ritchie D. A.; Shields, A. J. *Nature* **2006**, *439*, 179.

(6) Minot, E. D.; Kelkensberg, F.; van Kouwen, M.; van Dam, J. A.; Kouwenhoven, Leo P.; Zwiller, V.; Borgstrom, M. T.; Wunnicke, O.; Verheijen, M. A.; Bakkers, E. P. A. M. *Nano Lett.* **2007**, *7*, 367.

(7) Bimberg, D.; Grundmann, M.; Ledentsov, N. N. *Quantum Dot Heterostructures*; Wiley, New York, 1998.

(8) Zrenner, A.; Beham, E.; Stufler, S.; Findeis, F.; Bichler, M.; Abstreiter, G. *Nature* **2002**, *418*, 612.

(9) Blakesley, J. C.; See, P.; Shields, A. J.; Kardynał, B. E.; Atkinson, P.; Farrer, I.; Ritchie, D. A. *Phys. Rev. Lett.* **2005**, *94*, 067401.

(10) Santori, C.; Pelton, M.; Solomon, G.; Dale, Y.; Yamamoto, Y. *Phys. Rev. Lett.* **2001**, *86*, 1502.

(11) Reimann, S. M.; Manninen, M. *Rev. Mod. Phys.* **2002**, *74*, 1283.

(12) Fallahi, P.; Bleszynski, A. C.; Westervelt, R. M.; Huang, J.; Walls, J. D.; Heller, E. J.; Hanson, M.; Gossard, A. C. *Nano Lett.* **2005**, *5*, 223.

(13) Monat, C.; Alloing, B.; Zinoni, C.; Li, L. H.; Fiore, A. *Nano Lett.* **2006**, *6*, 1464.

(14) (a) Kouwenhoven, L. P.; Jauhar, S.; McCormick, K.; Dixon, D.; McEuen, P. L.; Nazarov, Yu. V.; van der Vaart, N. C.; Foxon, C. T.; *Phys. Rev. B* **1994** *50*, 2019. (b) Brandes, T.; Renzoni, F.; *Phys. Rev. Lett.* **2000** *85*, 4148.

(15) Alhassid, Y. *Rev. Mod. Phys.* **2000**, *72*, 895.

(16) Saito, T.; Schulman, J. N.; Arakawa, Y. *Phys. Rev. B* **1998**, *57*, 13016.

(17) Walther, C.; Blum, R. P.; Niehus, H.; Masselink, W. T.; Thamm, A. *Phys. Rev. B* **1999**, *60*, R13962.

(18) Maltezopoulos, T.; Bolz, A.; Meyer, C.; Heyn, C.; Hansen, W.; Morgenstern, M.; Wiesendanger, R. *Phys. Rev. Lett.* **2003**, *91*, 196804.

(19) Okada, Y.; Miyagi, M.; Akahane, K.; Kawabe, M.; Shigekawa, H. *J. Cryst. Growth* **2002**, *245*, 212.

(20) (a) Tanaka, I.; Kamiya, I.; Sakaki H.; Qureshi, N.; Allen, S. J. Jr.; Petroff, P. M. *Appl. Phys. Lett.* **1999**, *74*, 844. (b) Okada, Y.; Miyagi, M.; Akahane, K.; Iuchi, Y.; Kawabe, M. *J. Appl. Phys.* **2001**, *90*, 192.

(21) The incident light is introduced to the detection area from the bottom side of thinned sample to avoid tip scattering. Considering that the photon energy is lower than the bandgap energy of GaAs at 300 K by 54 meV, the interband absorption should mainly occur in wetting layer and the quantum dots; also, the absorption in GaAs through Franz-Keldysh effect can not be ignored since the build-in field will be larger than $10^5$ V/cm in the depletion region of *n*-GaAs surface. As a comparison, significantly larger photocurrent has been detected when photon energy is larger than the GaAs bandgap with similar excitation intensity.

(22) Jdira, L.; Liljeroth, P.; Stoffels, E.; Vanmaekelbergh, D.; Speller, S. *Phys. Rev. B* **2006,** *73*, 115305.

(23) Oshima, R.; Kurihara, N.; Shigekawa, H.; Okada, Y.; *Phys. E* **2004**, *21*, 414.

(24) The possibility of simultaneously containing two or more photoholes in one QD can be neglected according to the illumination intensity we've applied in this study.